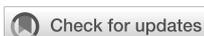

# Geomorphodynamics, evolution, and ecology of vertical roots

Martin Heidelman and Dervis Can Vural*

University of Notre Dame, Department of Physics, Notre Dame, IN, United States

The roots of some coastal and wetland trees grow peculiar vertical protrusions, the function of which remains unclear. Here, using computational simulations based on first-principles fluid and sedimentation dynamics, we argue that the protrusions work together to create an elevated patch of sediment downstream of the tree, thereby creating its own fertile flood-protected breeding grounds for the seedlings. In our simulations, we vary the vertical root diameter, root spacing and total root area and show that there is an optimal vertical root spacing that depends on root thickness. Next, we quantify and discuss the cooperative effects between adjacent vertical root patches. Lastly, by varying vertical root spacing of a patch of trees, we estimate a maximal vegetation density for which vertical-root production has a beneficial geomorphological response. Our hypothesis suggests that vertical roots, such as the 'knee roots' of baldcypress trees, have an important role in shaping riparian geomorphology and community structure.

KEYWORDS

regeneration mechanism, flood disturbance, riparian vegetation, cypress knees, mangrove roots, vertical roots, cone roots, pencil roots

## 1 Introduction

A number of tree species inhabiting coastal and wetland regions exhibit peculiar root structures that grow out of the soil vertically (see Figures 1A–C). Notable examples include the "knee roots" of Taxodium distichum (bald cypress), the "pencil roots" of the Avicennia genus and the "cone roots" produced by various species of Sonneratia and Bruguiera as well as Xylocarpus moluccensis Allen and Duke (2006); Srikanth et al. (2016).

Pencil roots and cone roots are common across mangroves, a class of species adapted to low-lying coastal areas typically of high salinity. A popular hypothesis is that these protrusions are pneumatophores, i.e. they act as snorkels that provide oxygen to the submerged parts of the root system Chomicki et al. (2014); Srikanth et al. (2016). This hypothesis is plausible due to the presence of specialized features, lenticels and aerenchyma, which allow gas to move into and throughout the plant when the vertical root is exposed Scholander et al. (1955); Armstrong (1979); Purnobasuki and Suzuki (2005); Seago et al. (2005); Arber (2010); Tomlinson (2016); Srikanth et al. (2016).

The function of another kind of vertical root, the "knees" of the bald cypress, is a bigger enigma (Figure 1A). These trees are long-lived conifers native to low-lying areas of southeastern United States and grow in a wide variety of soils along rivers, lakes, and swamps Shankman and Kortright (1994). In these frequently-flooded environments, seeds are primarily distributed by water Schneider and Sharitz (1988), where flood currents carry the seeds downstream and onto embankments or against protruding objects where they become embedded in sediment and germinate in the spring when flood levels have abated Streng et al. (1989); Huenneke and Sharitz (1990). The mechanism for knee formation was suggested to be the exposure of upper roots to the atmosphere during a flood drawdown. This metabolically favorable condition for respiration results in thicker layers of wood on the exposed region, which over time, creates a knee Whitford (1956). Various ideas have been proposed for why bald cypresses grow knee roots, including mechanical support Lamborn (1890), starch storage Brown (1984); Brown and Montz (1986), and gas exchange Kramer et al. (1952); Armstrong (1979); Briand et al. (2000); Martin and Francke (2015); Rogers (2021), however, neither of these explanations are entirely satisfactory. The mechanical hypothesis relies on the assumption that the knees provide support for a network of smaller roots extending downwards, however it was observed that these root networks were often not present Brown and Montz (1986). The starch storage hypothesis does not explain why trees growing in fluctuating wet and dry conditions need such an auxiliary organ and those that grow in constant wet or dry conditions do not.

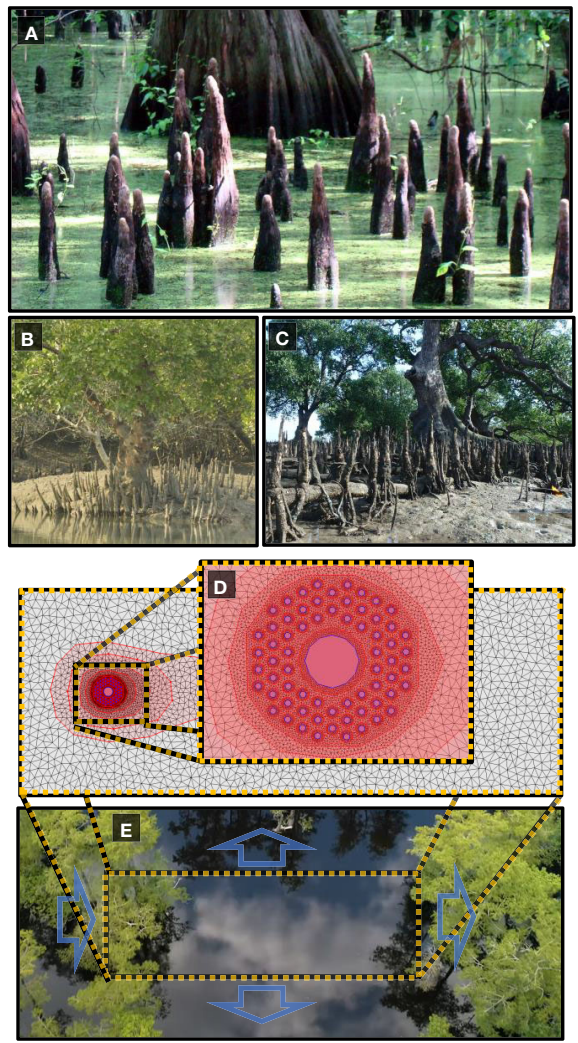

FIGURE 1
(A) Baldcypress knee roots (B) Cone roots of Xylocarpus moluccensis. (C) Cone roots of Sonneratia alba. (D) Vertical roots are modeled as circular obstacles in a two-dimensional unstructured mesh of varying resolution from 0.001 $m^2$ around each root to 1.3 $m^2$ far from the vegetation patch. (E) Trees exhibiting vertical roots are native to coastal and riparian environments subject to frequent flooding. Boundary conditions are set to mimic a vegetation patch with flood discharge from left to right with constant water height at each boundary edge. Images courtesy of (A) Jimmy Smith via Flickr available at https://www.flickr.com/photos/98937825@N00/2658124105. (B) Sagar Adhurya (C) Ria Tan "Perepat (Sonneratia alba)" by wildsingapore https://openverse.org/image/b0d46209-fd3c-4885-b057-b20aa91570df/ (E) Shawn Bannon via https://www.youtube.com/watch?v=JCuUpZXQDac.

The gas exchange hypothesis is not entirely satisfactory either. A field experiment concluded that due to the high rates of metabolic activity in cypress knees, the additional oxygen uptake was not great enough to significantly oxygenate the entire root system Kramer et al. (1952). Conversely, a laboratory experiment showed that cypress roots with knees exposed to the atmosphere indeed show increased oxygen levels Martin and Francke (2015), however the authors did leave open the possibility that knees are a response to another evolutionary driving force. The aeration hypothesis does little to explain the observation that cypress knees are not usually seen on trees growing in the most anoxic environments such as in perpetually deep water Kernell and Levy (1990); Briand et al. (2000); Martin and Francke (2015); Middleton (2020) and cypress knees do not show the characteristic aerenchyma and lenticels typical of pneumatophores Briand et al. (2000). Furthermore, if the knees indeed serve to exchange gasses, it is puzzling why the number of knees is inversely correlated with the maximum flood depth of the habitat Yamamoto (1992). Recently, a new but similar hypothesis proposes that rather than the knee's directly providing oxygen to the root system below, the knee's act as pumping stations, which due to their exposure to the atmosphere, are able to produce more sap in the emergent inner phloem tissue at the top of the knee which then is dispersed to the roots below Rogers (2021). But ultimately, with conflicting experimental evidence and perplexing field observations, it remains unclear if the knees are for aeration or pumping.

In this paper, we employ computer simulations based on physical first principles describing the fluid dynamics and geomorphodynamics surrounding the vertical roots, to compare the flow and elevation profiles in the vicinity of trees with and without vertical roots to better understand their function. In our simulations, we observe that the vertical root clusters shape the fluid flow around the tree, which in turn leads to substantial sediment accumulation downstream (called a "sand bar"), which we argue is their function / a function. We observe that a closely packed arrangement of vertical roots pulls the sand bar closer to the tree, which we discuss, could be preferential for certain life history strategies. By adding vertical roots in between arrays of trunks with varying density, we establish that vegetation growing at low densities benefit from knee production and vegetation growing at high densities are hindered by it.

We quantify how clusters of vertical roots modify the flow velocity and accumulate sediment, both as a function of the total area of the vertical root cluster, their spacing, and their thickness. We plot the soil erosion as a function of time at various distances away from a tree with vertical roots, and compare these numbers to that near a tree without vertical roots.

Additionally, we show that adjacent clusters of vertical roots operate cooperatively, in the sense that their collective benefit is more than the sum of their individual benefits; thus neighboring trees have an incentive to close the gap between themselves with vertical roots. The cooperative effect also predicts that trees growing in clusters would show significant advantages to isolated trees, thereby demonstrating the importance of spatial density in the dispersal of riparain vegetation. Interestingly, in our simulations, we find an optimal vertical root separation that maximizes the "viable soil area", i.e. the area of soil whose elevation is above a certain threshold. Moreover, we find the optimal root separation to be largely independent of the threshold chosen, and other physical system parameters, such as the water discharge rate. Remarkably, we find that the optimal value obtained from our simulations agrees with the empirically observed knee separation in bald cypress forests, which supports our hypothesis that vertical roots function as soil collectors.

Lastly, we should emphasize that neither of the available hypotheses (gas exchange, carbon storage, structural stability) explain the observation that baldcypress trees only grow knee roots in environments with fluctuating water levels (where there is flow) and not in persistently dry or submerged conditions, (where there is no flow). However, the hypothesis presented here does.

The sediment accumulation itself might serve multiple purposes, such as (1) the generation of elevated pioneer landforms which increase the ability for local vegetation recruitment by reducing the chance for a seedling to drown Naiman et al. (1993); Ward et al. (1999); Gurnell and Petts (2002); Gurnell (2014); Yagci and Strom (2022). As an additional synergistic effect, the vertical roots (2) slow down the water right at the elevated region downstream, thereby increasing the probability that seeds germinate there both by directing seeds in the downstream direction and stabilizing the elevated region Danvind and Nilsson (1997); Ashworth et al. (2000). There may be additional (possibly less-significant) benefits of the added sediment, such as (3) to better anchor the tree against fast winds (highly common in this habitat), and possibly, (4) to increase the soil nutrition available to both the parent and saplings through the accumulation of fine silts and organic particles Ward et al. (1999); Ashworth et al. (2000); Gurnell and Petts (2002).

As we conclude our introduction, we should emphasize that the implications of our simulations are not exclusive to protrusions that grow upwards from the soil, but also for more typical root structures growing from the trunk or branches into the soil, such as prop and stilt roots. Aerial roots that otherwise help the plant to climb and spread would also have a similar effect. As such, the generic term "vertical root" (and occasionally, the inaccurate shorthand, "knee") should throughout be understood as any structure above the soil line that is perpendicular to the direction of fluid flow.

## 2 Materials and methods

While similar numerical investigations exist in the literature, an analysis of the variables influencing the preservation of downstream sediment elevation is lacking. As such, one of the goals of our model is to quantify the downstream sediment patterns after a flood event for a variety of different vertical root configurations. To capture all of the potentially relevant fluid mixing within the 'root array', we chose to model the roots *via* simplified geometrical elements, rather than a porous media model Yamasaki et al. (2021b). Therefore, to incorporate a complex bed geometry into our computational mesh, we utilized the *MFlow*02 solver available through the International River Interface Cooperative (iRIC) Gamou (2011); Nelson et al. (2016); Shimizu et al. (2020). *MFlow*02 solves two-dimensional unsteady flow and riverbed variation and due to its use of an unstructured mesh, is practical for the study of geomorphodynamic evolution of complex bed surfaces including the development of sandbars at river confluences, and the geomorphological impact assessment of piers and vegetation Gamou (2011); Nones et al. (2018); Nones (2019); Ali et al. (2019); Shimizu et al. (2020); Liu et al. (2020); Yan et al. (2022). A shortcoming of our study is the use of this two-dimensional flow modeling package, which is likely to over predict the onset of turbulence due to the inverse energy cascade phenomenon of 2-dimensional turbulence Boffetta et al. (2012); Pouquet et al. (2013). Despite this shortcoming, our results agree well with previous experimental and numerical investigations with respect to the position and size of turbulent structures, such as the formation of the Von Karman Vortex Street Garcia (2020). For example, comparison between our model results and the numerical results presented in Nicolle and Eames (2011), gives a 7% difference in the length between the trailing edge of the root array and the formation of the Von Karman Vortex Street for similar model parameters.

### 2.1 Simulation details

Vertical roots are modeled as circular obstacles within a two-dimensional rectangular domain with an optimized irregular mesh. The mesh was created with the mesh generation tool in the iRIC solver suite. Mesh convergence tests were performed by comparing the sizes of hydrological features in the wake region to those previously published in order to assure model accuracy while minimizing computational cost. The mesh generation tool of the iRIC allows mesh refinement regions to be included allowing much finer mesh resolution around each vertical root than regions far from the root array. After convergence tests to within a 7 percent difference with similar models, the maximum area for mesh elements around each vertical root was set to $0.0006m^2$ for 10 centimeter and 14 centimeter diameter roots and $0.001m^2$ for roots 20 centimeters in diameter or larger (see Figure 1 panels d and e for a visual example of the two-dimensional model). The maximum area of mesh elements is then gradually increased to $1.3m^2$ far from the root array. Minimum angle between cell vertices is set to 20 degrees. This results in the total number of elements in each model to range between 11,000 and 18,000 depending on the number of root refinement regions being used.

For most simulations, root diameter was kept constant at 20cm (root diameters are varied between 10cm and 80cm) and are evenly spaced at the vertices of an equilateral hexagonal array. Initial bed elevation was set to ten meters across the entire model area, and initial water depth is set to one meter. Boundary conditions were selected to simulate a typical hydrologic regime that would be experienced by a riparian tree in a floodplain forest at peak flood levels as depicted in Figure 1E). The blue arrows represent the boundary conditions on each of the four sides of the model. Flood discharge from the left to the right side of the model area was varied between $40m^3/sec$ to $10m^3/sec$. Boundary conditions on all other three sides were set to a constant water height of 1 meter, allowing both sediment and water movement across the boundary. To achieve results applicable to a variety of soil types, we used a mixed grain diameter sediment model with a 'well-mixed' sediment profile consisting of particles in the size range from 0.001mm to 34mm and a $D_{50}$ of 3.2mm. The morphological factor was kept at 1, relative weight of bedload material was set to 1.65, void ratio of bed material was 30 percent, and the exchange layer thickness was 0.5m. Flow was allowed to stabilize for 500 seconds to avoid any effects from initial flood wave motion. Computational time steps ranged from 0.01 seconds to 0.08 seconds depending on the density of the root array and flood time was kept constant at 25000 seconds for a total of 24500 seconds of morphological change. Flood time was determined by allowing the discharge event to continue until the bed elevation reached a steady state.

## 2.2 Model equations

In MFlow02, water continuity is described by

$$\partial_t h + \nabla \cdot (Vh) = 0 \qquad (1)$$

where $t$ is time, $V$ is the 2-D flow velocity vector, and $h$ is the water depth. The equations describing the momentum of the flow are,

$$D_t(uh) - f(uh) = -gh\,\partial_x H + D^{vy}(uh) - \tau_{bx}/\rho$$

$$D_t(vh) + f(vh) = -gh\,\partial_y H + D^{vy}(vh) - \tau_{bx}/\rho$$

Here, $D_t \equiv \frac{\partial}{\partial t} + V \cdot \nabla$ is the material derivative, $D^{vy} = \partial_x(v\,\partial_x) + \partial_y(v\,\partial_y)$, $u$ and $v$ are the flow velocity components in the $x$ and $y$ direction, $v$ is the kinematic eddy viscosity, $g$ is the gravitational acceleration, $H$ is the water surface level (depth + ground elevation), $f$ is a Coriolis parameter, $\rho$ is the water density, and $\tau_{bx}, \tau_{by}$ are the bottom shear stress component in the $x$ and $y$ direction given by $\tau_{bx}/\rho = C_f u\sqrt{u^2+v^2}$ and $\tau_{by}/\rho = C_f v\sqrt{u^2+v^2}$. $C_f$ is the riverbed friction coefficient given by $C_f = gn^2/h^{1/3}$. The kinematic eddy viscosity $v$, is calculated from the standard k-e model as, $v = C_\mu \kappa/\epsilon$ where $\epsilon = C_e \kappa^{3/2}/l$ is the energy dissipation rate, l is the length scale of the turbulence, $C_\mu$ is a constant equal to 0.09, $\kappa$ is the turbulent energy given by, $\kappa = 2.07 u_*^2$, where $u_*$ is the bottom friction velocity given by, $u_* = n\sqrt{g(u^2+v^2)}/h^{1/6}$, where, n is the Manning roughness coefficient. and The dimensionless shear stress, $\tau_*$ used to calculate sediment discharge is calculated as, $\tau_* = u_*^2/sgd$, where s is the submerged specific gravity of the suspended sediment particle and d is the diameter of the particle. Bed load sediment discharge is calculated from the Meyer-Peter-Muller formula,

$$q_b = 8\sqrt{(\sigma/\rho - 1)gd^3}(\tau'_* - \tau_{*c})^{3/2} \qquad (2)$$

where $d$ is the sediment grain diameter, $\sigma$, is the gravel density Meyer-Peter and Muller (1948). $\tau_{*c}$ is the critical tractive force calculated from the Iwagaki formula Iwagaki (1956). $\tau'_*$ is calculated from the Kishi and Itakura formula Itakura and Kishi (1980). The total sediment discharge set by the formula above is divided into a sediment discharge in both the normal (n) and tangential (s) direction of the river flow streamline as, $q_i = q_b(v_{bi}/V_b - \kappa\frac{\partial z}{\partial i})$ where $i$ stands for either the n or s direction, $\kappa = \sqrt{\tau_{*c}/\mu_s\mu_k\tau_*}$, $\mu_s$ is the static friction factor, $\mu_k$ is the kinetic friction factor, $z$ is the height of the river bed, and $v_{bi}$ is $v_{bs}$ and $u_{bn}$ and are the (s) and (n) directional components of the flow velocity near the river bed and are calculated as, $v_b = 8.5 u_*$ and $u_b = -7 v_b h/r$. $V_b$ is the absolute value of the flow velocity near the river bed. Here $r$ is the radius of curvature of the river flow streamline calculated as,

$$\frac{1}{r} = \frac{1}{(u^2+v^2)^{3/2}}[u(u\frac{\partial v}{\partial x} - v\frac{\partial u}{\partial x}) + v(v\frac{\partial v}{\partial y} - v\frac{\partial u}{\partial y})] \qquad (3)$$

The velocity of buoyancy of suspended sediment is calculated from the Itakura-Kishi formula Itakura and Kishi (1980), $E_s = K[\alpha_* \rho u_* \Omega/(\tau_* \sigma w_0) - 1]$ where, $K$ is a constant equal to 0.008, $\alpha_*$ is a constant equal to 0.14, and $w_0$ is the sedimentation speed calculated from the Rubey formula. The concentration of suspended sediment at the referential level can be calculated using the Lane-Kalinske formula Lane and Kalinske (1941),

$$C/C_\alpha = \exp\left(-6\frac{w_0}{\kappa u_*}\frac{z - z_\alpha}{h}\right) \qquad (4)$$

where, $C$ is the concentration of suspended sediment, $z_\alpha$ is a referential height equal to $0.05h$. The conservation of mass of the depth-averaged concentration of suspended sediment is described by,

$$D_t(ch) = (E_s - C_\alpha)w_0 + \frac{\partial}{\partial x}D_x\frac{\partial(ch)}{\partial x} + \frac{\partial}{\partial y}D_y\frac{\partial(ch)}{\partial y}$$

where, $D_x$ and $D_y$ are the diffusion coefficients of the suspended sediment. The change in elevation, z is then,

$$\frac{\partial z}{\partial t} + \frac{\partial q_x}{\partial x} + \frac{\partial q_y}{\partial y} - (E_s - C_\alpha)w_0 = 0 \qquad (5)$$

Where, $q_x = q_n\cos(\theta) - q_s\sin(\theta)$ and $q_y = q_n\cos(\theta) + q_s\sin(\theta)$ and $\theta = \tan^{-1}(v/u)$.

A mixed grain diameter model was run, where the accumulation curve of riverbed grains is divided into n hierarchies. A representative grain diameter $d_k$ indicates the existence possibility $p_k$ of a particular representative grain. Central grain diameter is defined as, $d_m = \sum_{k=1}^{n} p_k d_k$. Additionally, the sheltering effect when calculating the non-dimensional critical tractive force of each grain diameter is calculated as,

$$\frac{\partial \tau_{ck}}{\partial \tau_{cm}} = \begin{cases} 0.85 & \text{if} \quad d_k/d_m \le 0.4 \\ \frac{d_k/d_m}{1+0.34\log\ (d_k/d_m)} & \text{if} \quad d_k/d_m > 0.4 \end{cases}. \tag{6}$$

The conservation of sediment volume is then expressed as,

$$\frac{\partial E_b}{\partial t} + \frac{\partial z}{\partial t} + \frac{\partial}{\partial x}\sum_{k=1}^{n}p_k q_{xk} + \frac{\partial}{\partial y}\sum_{k=1}^{n}p_k q_{yk} = \sum_{k=1}^{n}(E_{sk} - C_{ak})w_{0k}$$

# 3 Results

Our solver obtains both the fluid velocity field and the resulting force governing the sediment motion (cf. Materials and Methods). First, we compare the differences in fluid flow and downstream sediment motion around a tree with and without vertical roots. Then, by making variations in vertical root density in the cluster, we determine the optimal root density that would maximize the downstream area above a chosen threshold. We then quantify the peak position of the sediment accumulated downstream, a variable that is important for certain plant life histories. Then, by using vegetation clusters containing two different sizes, we determine the density of vegetation that would benefit from growing knees. Last, we investigate the cooperative effects between adjacent vertical root patches,

## 3.1 Vertical roots reduce erosion

First we compare the sediment height (Figures 2A, C) and fluid velocity profile (Figures 2B, D) around a tree with vertical roots with one without. When a tree is surrounded by vertical roots, a sediment bar of approximately one meter above the surrounding topography forms between one and eight patch lengths downstream of the tree. When the tree has no vertical roots, no downstream bar is formed. The elevation versus time plot in Figure 2 shows the elevation at 1, 4, and 8 patch lengths downstream for the tree with knees. Only a single curve is given for the tree without knees as no spatial variation in sediment height was observed.

The difference in downstream elevation can be explained by the increased drag on the fluid causing a delay in patch-scale turbulence formation due to bleed flow through the cylinder array. A more detailed description of this wake hydrology can be found in Yagci et al. (2016); Chang et al. (2017); Kitsikoudis et al. (2020). This consequence is particularly advantageous for plants whose seeds require deposition onto an exposed surface. As such, vertical roots can be viewed as organs that generate downstream protected microhabitats where the seeds can successfully germinate.

## 3.2 An optimal vertical root density minimizes sediment loss

To determine how knee density (number per area, $\rho=N/A$) affects downstream bar production, we position equally sized knees (20cm in diameter) at equal distances from one another at the vertices of an equilateral hexagonal grid. We vary one of the three variables, $\rho$, $A$, $N$, and observing another, while keeping the third variable constant. In Figure 3, panels a and d are the total downstream area above three different threshold depths, and the percent velocity change respectively, for trials holding the patch area constant and varying the density. In panels b and e, we plot the same two quantities, but this time, holding the number of roots constant and varying their separation. Similarly, c and f vary the number of roots, while holding root spacing constant. In Figure 3A, a maximum is observed where an increasing vertical root density stops producing more viable downstream area. This point will be referred to as the optimal root density.

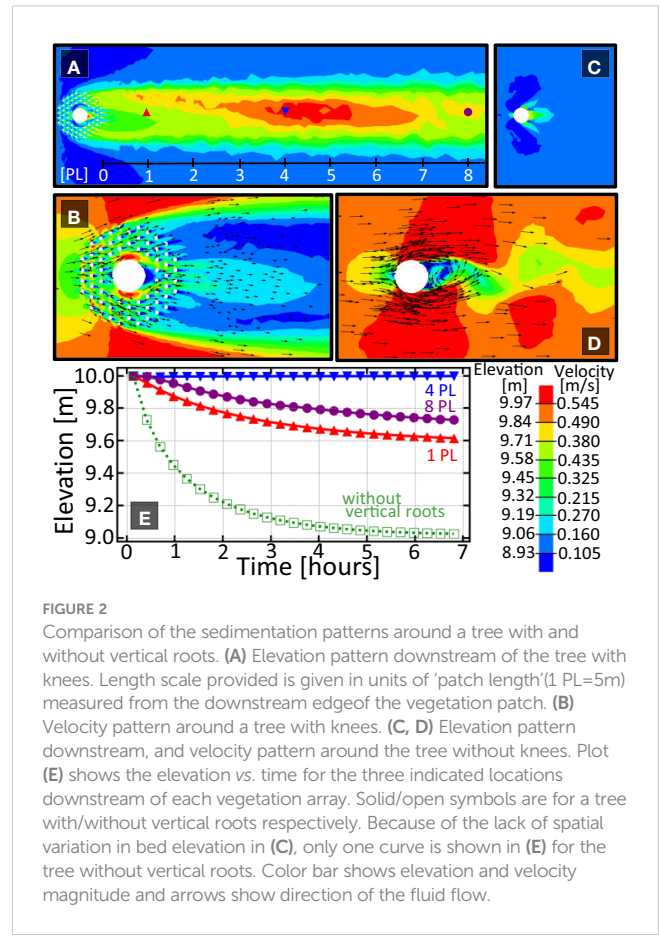

FIGURE 2
Comparison of the sedimentation patterns around a tree with and without vertical roots. (A) Elevation pattern downstream of the tree with knees. Length scale provided is given in units of 'patch length'(1 PL=5m) measured from the downstream edgeof the vegetation patch. (B) Velocity pattern around a tree with knees. (C, D) Elevation pattern downstream, and velocity pattern around the tree without knees. Plot (E) shows the elevation vs. time for the three indicated locations downstream of each vegetation array. Solid/open symbols are for a tree with/without vertical roots respectively. Because of the lack of spatial variation in bed elevation in (C), only one curve is shown in (E) for the tree without vertical roots. Color bar shows velocity magnitude and arrows show direction of the fluid flow.

In previous studies, clump-type vegetation has been modeled as regularly spaced cylinders described by the solid volume fraction, $\phi = N(d/D)^2$ where $N$ is the number of cylinders, $d$ is the diameter of a cylinder and $D$ is the diameter of the circular patch Tanaka and Yagisawa (2010); Nicolle and Eames (2011); Chang and Constantinescu (2015); Chang et al. (2017). For comparison purposes, we also provide the $\phi$ values of all the arrangements.

As we see in Figure 3, downstream elevation is best preserved at densities between three and four roots per $m^2$($\phi$ between 0.12 - 0.14). Percent velocity change was calculated from the ratio of average fluid velocity directly downstream of the root array to the average fluid velocity far upstream of the root cluster. In Figure 4 we show the contour plots of the downstream elevation and fluid velocity around each of the three root clusters denoted by roman numerals in panel d) of Figure 3. We show in Figure 3 that as patch density increases, the percent velocity change for the fluid entering and leaving the root array also increases. For densities lower than the optimal density, fluid leaving the root array is slowed, acting as a wake barrier and preventing the re-connection of the faster outer streamlines. This decreased velocity leads to less erosion and increased sedimentation, resulting in an elevated downstream bar. When the optimal density is reached, the root array begins to generate downstream patch-scale turbulence causing a truncation of the sedimentation region and bringing the elevated region nearer to the root array. As seen in panels c and f, the solid volume fraction $\phi$ is not always a good predictor of downstream area preservation. The diameter of the patch must also be large enough to separate the flow long enough to create a bar.

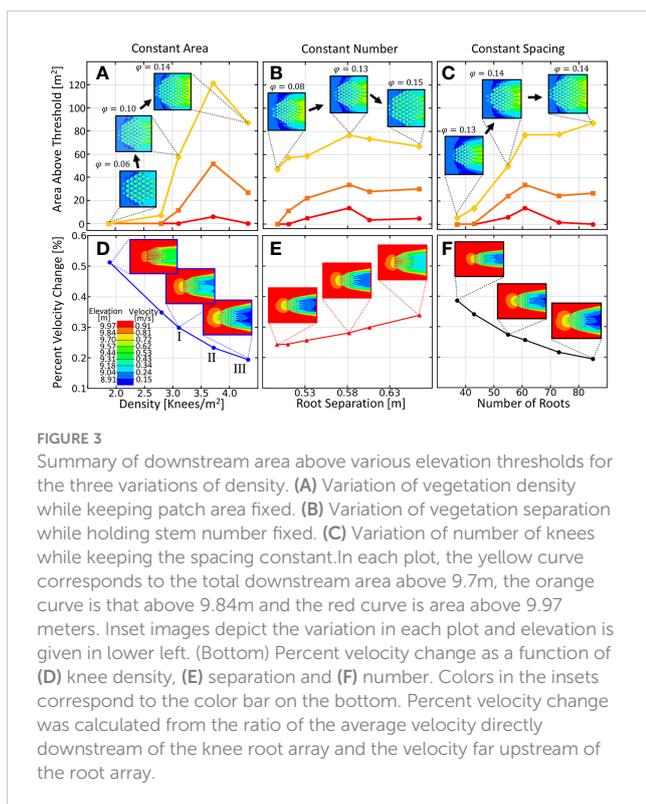

Another interesting feature of our simulation results depicted in Figure 4 is that, further increasing the root density beyond the optima decreases the distance between the root array and the elevated peak. This suggests that trees with less-dispersive propagules would be selected for higher root density as to draw the bars closer, rather than simply maximizing the downstream area for their seedlings.

Species whose vertical roots grow continuously (such as the baldcypress), would start out with a lower root density, encouraging seedling growth farther away from themselves; however, as the tree ages and its knees grow larger and wider, the downstream bar will move closer to itself, ultimately leading to a seedling replacing its parent.

In Figure 4, we observe a transition in the velocity contour plots from a slowed steady wake into patch scale turbulence, which occurs when root density is increased. This trend is consistent with the wake structures experimentally observed in Tanaka and Yagisawa (2010); Chen et al. (2012); Zong and Nepf (2012). Furthermore, the resulting increase in downstream elevation observed in Figure 4 agree with the trend empirically shown in Tanaka and Yagisawa (2010), except, in our simulations, we are able to observe that the total viable soil area stops increasing once the vertical root density hits an optimum value.

Nicolle and Eames (2011) observed that patch-scale turbulence occurred at $\phi$ values between 0.0884 and 0.1451. The optimal root density that we find here (Figure 3) lies in the same regime. Also, our observed pattern in wake characteristics with increasing $\phi$ agrees qualitatively with the fully three-dimensional large eddy simulations of Chang and Constantinescu (2015). However, our finer sampling of root density as well as our more detailed description of the downstream elevation profile reveals an optimal peak in soil area as well as the dependence of bar position on root density.

The different colored curves in Figures 3A–C, depict the results for when a different threshold is used during analysis. The red curve depicts the amount of area downstream of the root array that is above a threshold of 9.97m, the orange curve for a threshold of 9.84m and the yellow curve for a threshold of 9.7m. As we see, remarkably, the threshold value does not change the location of the optimal point on the curve.

## 3.3 Vertical roots are only beneficial for low density vegetation

The root arrays in the previous simulations could just as well be thought of as individual vegetation stems, indicating that an optimal vegetation density for maximal downstream propagation success exists. Vegetation with canopy size comparable to its stem diameter or that have shade tolerance might be able to reach such optimal densities. However, vegetation with large canopies must grow at lower densities and thus would need to produce vertical roots to obtain maximal downstream reproductive success. To determine the cutoff of when it is beneficial for vegetation to produce vertical roots, a series of additional simulations were run where smaller diameter knee roots are added into larger arrays of tree trunks. Tree trunks are modeled as either 40 or 80cm diameter obstacles in the computational mesh and the knee roots are modeled as 10 cm diameter obstacles. For each given trunk configuration, knee roots are added and the area above a 9.71 m threshold is calculated. These data can be seen in Figure 5 as the open symbols and dashed lines. For configurations that are already of significant solid volume fraction, any additional knees results in a decrease in downstream area above the threshold. However, configurations of low density see a large benefit in producing knees. This shows that vegetation growing below the optimal solid volume fraction for their diameter will see a benefit in growing knees, and vegetation able to grow at or above this density would see a decreasing benefit.

## 3.4 Adjacent root patches cooperate

Our next set of simulations quantify the cooperativity between neighboring trees, as mediated by their vertical roots. In these simulations, two root patches consisting of 37 twenty centimeter diameter roots at their optimal separation were spaced at varying distances from one another.

The root patches were offset so that the edges of the patches aligned. Figure 6 (left) depicts the downstream area above a 9.71 m threshold for these two adjacent root patches as a function of the distance between the two patch centers. Contour plots of the downstream elevation profile for the labeled data points as well as that of a single patch are given in Figure 6. An increasing cooperative advantage is seen as the patch distance decreases, indicating that trees should seek to close the gap between themselves and their neighbors to reap the maximal cooperative effect. However, a gap up to 3 m will still provide additional cooperative effects. Our theoretical results agree with the experimental flow tank observations, in that the wake interactions between neighboring patches enhance sand bar growth Yamasaki et al. (2021a), and that enhanced downstream deposition is observed for two closely neighboring patches Meire et al. (2014).

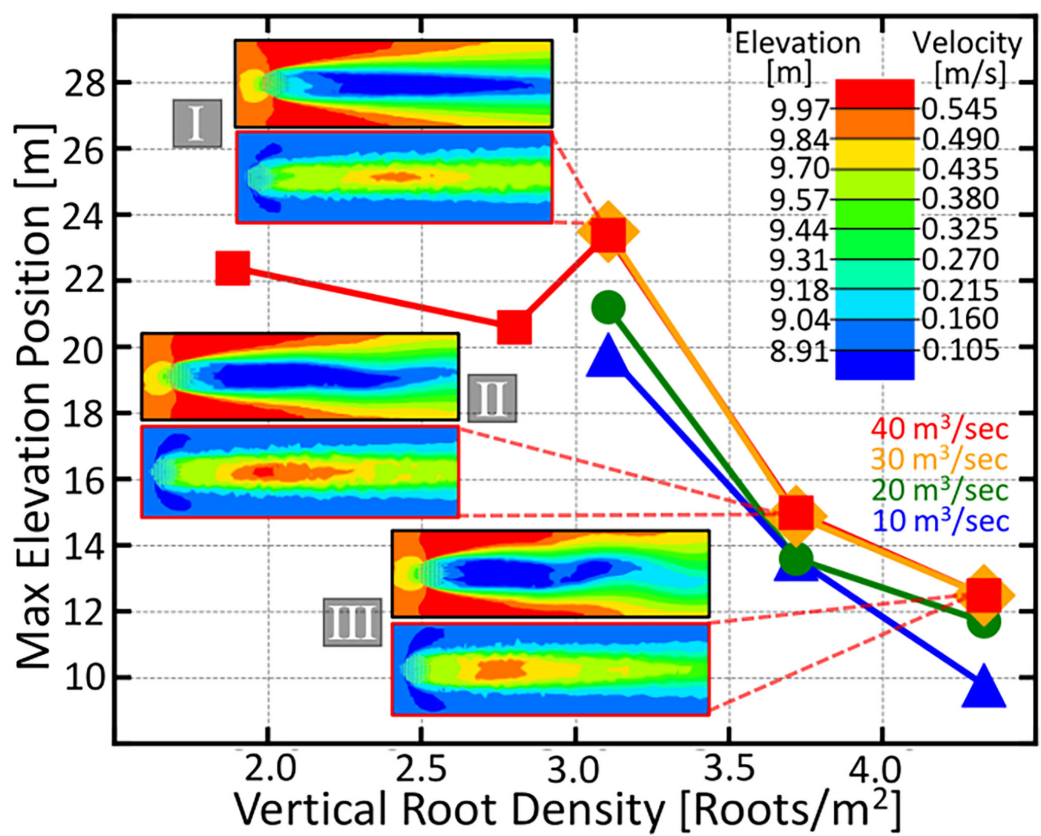

FIGURE 4
The maximum elevation position downstream from the root patch as a function of vertical root density. Contour plots depicting the downstream elevation after the flood event and the fluid velocity around and downstream of the four knee root arrays marked with roman numerals in Figure 3 are given. Velocity magnitude and elevation are given by the color scale in the upper right. The four color coded curves represent different flood discharges ranging from 10 $m^3$ / secto 40 $m^3$ / sec and are given by the labels in the lower right. The greatest densities have the closest elevated bar region, a consequence that could be advantageous for a life history which requires an adjacent elevated microsite for establishment.

In Figure 6 (right), we show how clusters of two and three trees with knees work together to produce considerably more viable area than a single tree. In addition to cooperating with neighboring trees, the vertical roots of an individual tree may also benefit from this cooperativity effect. The inset image in the lower right of Figure 6 shows a knee grouping, which may be a self-cooperating strategy to reduce biomass expenditure while maintaining large sedimentation effects.

Another pattern seen in Figure 6 is that the root patch furthest downstream reaps a greater benefit from the cooperative effect due to sandbar location. Future simulations with more vegetation patches in different arrangement angles could reveal further interesting competitive and cooperative dynamics between multiple root patches.

# 4 Discussion

## 4.1 Evolutionary and ecological implications

Figure 4 suggests that a tree can modify its geomorphological influence to suit its life history by altering the density of its vertical roots. By increasing the root density beyond 3.7 roots /$m^2$, a tree can draw its downstream bar towards itself, which would be more advantageous for trees whose propagules are adapted for close germination, such as bruguiera gymnorhiza which produces cigar shaped propagules that are often viviparous and germinate after sticking into the mud. In contrast, trees whose propagules more readily float and dispersed hydrochorically such as Xylocarpus moluccensis and Sonneratia alba would perhaps benefit from maximizing the overall area suitable for germination. Long lived species like Taxodium distichium may reap both rewards, where early in their life when their knees are of smaller diameter and the vertical root patch is at a lower solid volume fraction, offspring will germinate farther away from the parent. As the tree ages and the knees become larger and the solid volume fraction increases, the downstream bar will move closer towards itself, where one of its offspring can eventually overtake the parent's spot in the forest.

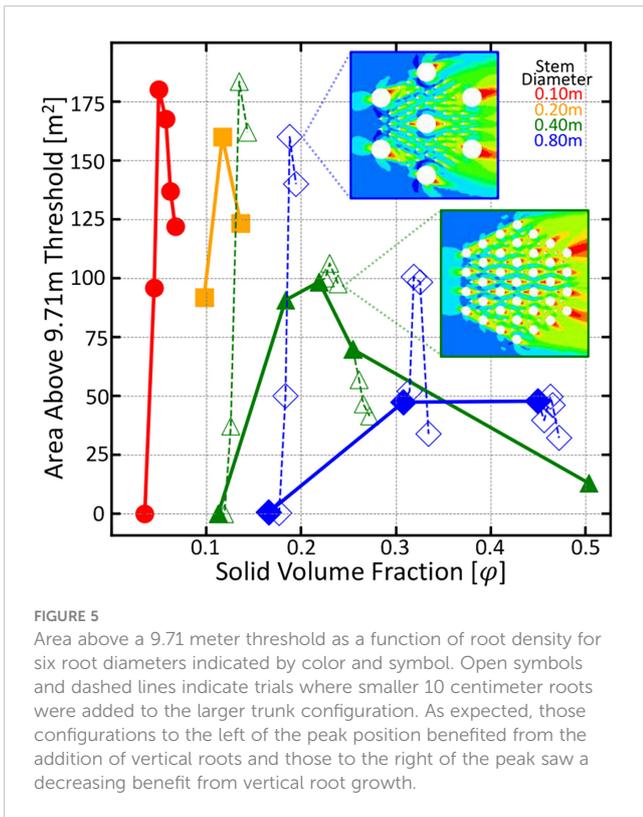

FIGURE 5
Area above a 9.71 meter threshold as a function of root density for six root diameters indicated by color and symbol. Open symbols and dashed lines indicate trials where smaller 10 centimeter roots were added to the larger trunk configuration. As expected, those configurations to the left of the peak position benefited from the addition of vertical roots and those to the right of the peak saw a decreasing benefit from vertical root growth.

Our results on the cooperativity between two root patches emphasizes the advantage of having neighbors who also produce vertical roots. In cypress forests, the fluctuating recruitment rate of seedlings often leads to patches of similar aged trees. After an initial competition for access to sunlight, the surviving neighboring trees turn into allies as their combined footprint of knees enhances the amount of fertile germination grounds for their seeds.

Both the aeration and sedimentation properties of vertical roots could be the reason these appendages were selected for in wetland environments. In mangroves, the presence of lenticels and aerenchyma suggest a closer tie to gas exchange, while in baldcypress the lack of these structures suggests a closer tie to sedimentation. In their 150 million years of existence (Taylor and Taylor, 1993; Seward, 2010) baldcypress trees have endured intense flooding events, like those at the end of the quaternary ice age, possibly because of their vertical roots that allowed them to spread into flood scoured plains when other species could not.

It is interesting to note that their close relative, the pond cypress (Taxodium ascendens) appears to be losing their reliance on the production of knees and also have a tendency to grow to smaller dimensions. It could be the case that this recent speciation is a response to less severe flood patterns. By growing at greater densities they instead alter geomorphology to their advantage at a group (rather than individual) scale, consistent with our simulations and their interpretation, as discussed in the subsection "Vertical roots are only beneficial for low density vegetation". It would be interesting to test this idea empirically: Trees with greater canopy spreads would need to grow at a lower densities and therefore would be in a greater need of vertical roots, so if our claims regarding knee function is correct, then cypress knee root production should correlate with the canopy spread of the tree.

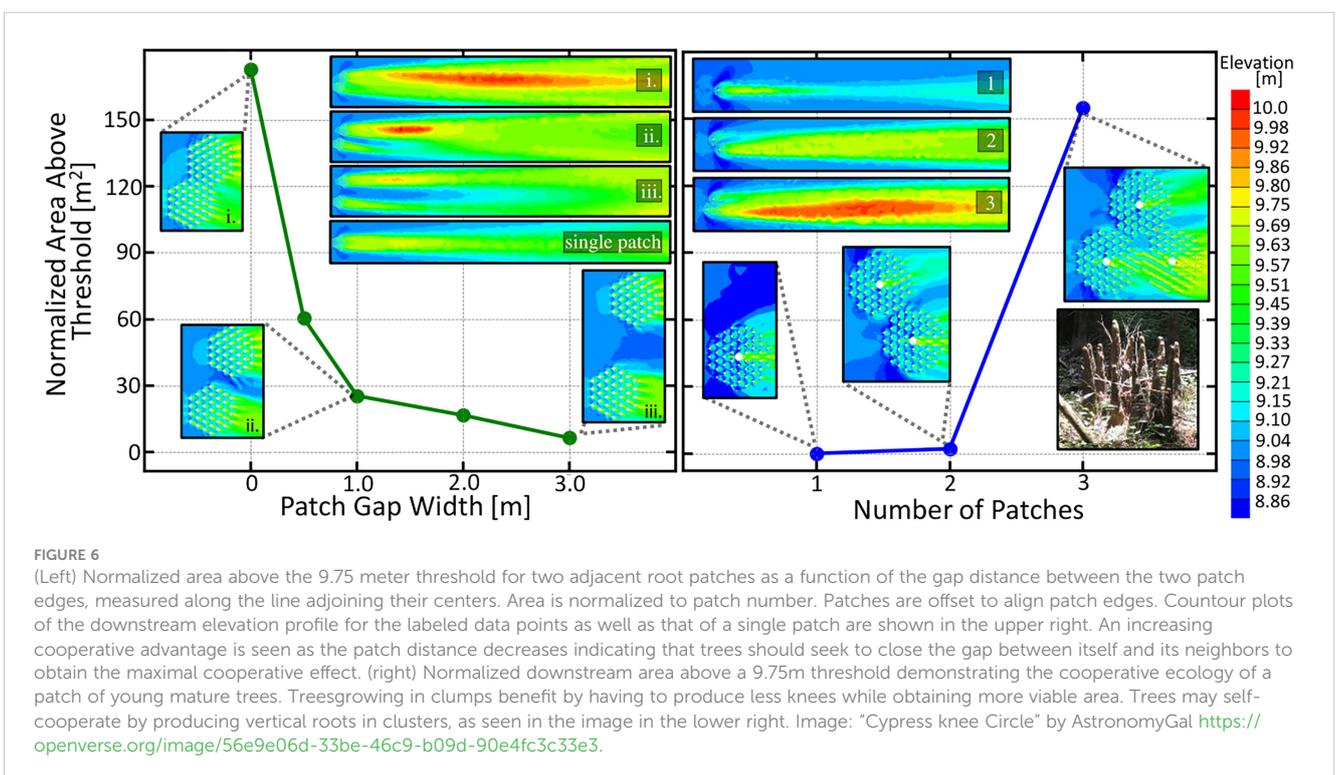

FIGURE 6
(Left) Normalized area above the 9.75 meter threshold for two adjacent root patches as a function of the gap distance between the two patch edges, measured along the line adjoining their centers. Area is normalized to patch number. Patches are offset to align patch edges. Contour plots of the downstream elevation profile for the labeled data points as well as that of a single patch are shown in the upper right. An increasing cooperative advantage is seen as the patch distance decreases indicating that trees should seek to close the gap between itself and its neighbors to obtain the maximal cooperative effect. (right) Normalized downstream area above a 9.75m threshold demonstrating the cooperative ecology of a patch of young mature trees. Trees growing in clumps benefit by having to produce less knees while obtaining more viable area. Trees may self-cooperate by producing vertical roots in clusters, as seen in the image in the lower right. Image: "Cypress knee Circle" by AstronomyGal https://openverse.org/image/56e9e06d-33be-46c9-b09d-90e4fc3c33e3.

## 4.2 Empirical support

The outcomes of this paper are consistent with a number of qualitative and quantitative earlier observations. For example, it was shown earlier that cypress knees lessen the storm runoff capabilities of streams Miroslaw-Swiatek and Amatya (2017) and since baldcypress seeds are most likely to germinate in regions of sediment exposed during flood drawdown Demaree (1932); Titus (1990); Rutherford (2015), an elevated downstream region would act as a prime location for its seeds to germinate. Additionally, baldcypress seedling mortality was tied to the elevation at which they germinate, with seedling death occurring after 60 days of submergence Conner et al. (1986); Souther and Shaffer (2000); Middleton (2000). Here, we showed that a tree with knees is much more effective at producing these downstream protected microsites than a tree without.

The role of vertical mangrove roots on sediment accumulation has also been noted in qualitative empirical observations, particularly in their value in coastline protection Feller et al. (2010); Lee et al. (2014). Interestingly, vertical roots (the pneumatophores) were better at conserving sediment in mangrove forests when compared to stilt roots Krauss et al. (2003). Additionally the amount of oxygen uptake was found to be less in their cone roots than in the stilt roots Kitaya et al. (2002), strengthening our hypothesis.

The bidirectional feedback loops between vegetation coverage and the geomorphodynamics of regions with high hydrodynamic influence is lately gaining increased attention Cornacchia et al. (2019); Larsen (2019); Yamasaki et al. (2021a); Huai et al. (2021) with vegetation found to self-organize itself in environments of hydrologic influence Schwarz et al. (2018); Cornacchia et al. (2020). Vegetation density has been determined to be an important factor in how these ecosystems develop and has led to a range of experimental and computational work on understanding how clump-type vegetation affects hydraulic processes Tanaka and Yagisawa (2010); Chen et al. (2012); Zong and Nepf (2012); Meire et al. (2014). Our findings are consistent with these previous studies and expand upon them by looking more closely at the parameter space which affects downstream sediment bar size and shape.

More recently, flume tank experiments have investigated the drag coefficients of various vegetation arrays Nair et al. (2022), and looked at quantifying the incipient motion of sediment, suggesting an optimal vegetation patch porosity to reduce erosion Kazemi et al. (2021). Because the focus of this study was on the incipient motion of downstream sediment, they gathered data only for approximately one patch length downstream, and used only three sizes of 9 cylinders which provides a porosity resolution too wide to adequately describe the downstream sedimentation characteristics in the flow regime where patch scale turbulence is induced.

Our results suggest that the primary adaptive advantage of vertical roots is to constitute a clump-type vegetative profile to maximize its impact on the fluid flow. This observation deepens our understanding of the vegetation and geomorphology feedback loop. It demonstrates that vegetation does not simply play a passive role in influencing flow patterns, but rather puts energy resources into harnessing geomorphodynamic processes to its own benefit.

## 5 Conclusion

We have shown that vertical roots can be an advantageous adaptation for species which rely on exposed soil for germination and benefit from elevated ground to minimize the risk of drowning. This is particularly plausible for cypress knees, whose function has remained an open question.

Our data suggests that the knees influence sedimentation and fluid flow to their advantage, emphasizing that vegetation does not always play a passive role in flood hydrodynamics. We propose that cypress knees are a flood adaptation, produced when the tree experiences erosion around the base. Exposure of the upper roots to oxygen during flood draw down provides the metabolic resources needed for rapid growth and explains why knees are mostly seen in habitats with fluctuating wet and dry conditions. We further have shown that the cooperative effects between neighboring trees are significant and that vertical roots can play a crucial function for the downstream spread of pioneer plant species. Finally, we have shown that vertical roots are only evolutionarily advantageous to vegetation that must grow at small densities such as large woody vegetation.

## Data availability statement

The original contributions presented in the study are included in the article/supplementary material. Further inquiries can be directed to the corresponding author.

## Author contributions

MH and DV conceived idea and wrote the manuscript. MH implemented computational methodology, compiled results, and composed figures. DV advised and aided throughout the process. All authors contributed to the article and approved the submitted version.

## Conflict of interest

The authors declare that the research was conducted in the absence of any commercial or financial relationships that could be construed as a potential conflict of interest.

## Publisher's note

All claims expressed in this article are solely those of the authors and do not necessarily represent those of their affiliated organizations, or those of the publisher, the editors and the reviewers. Any product that may be evaluated in this article, or claim that may be made by its manufacturer, is not guaranteed or endorsed by the publisher.